\documentclass[fleqn,10pt]{wlscirep}
\usepackage[utf8]{inputenc}
\usepackage[T1]{fontenc}
\usepackage{bm}
\usepackage{xcolor}
\definecolor{v}{rgb}{0.6, 0.2, 0.8} 
\usepackage{xcolor}
\usepackage{enumerate}

\title{Gravitational friction from d'Alembert's principle}

\author[1,*]{C. Ortiz}
\author[2]{R.S. Khatiwada}

\affil[1]{Universidad Autónoma de Zacatecas, Unidad Académica de Física, Zacatecas, 98060, México}
\affil[2]{Golden Gate International College, Tribhuvan University, Kathmandu, 446007, Nepal}

\affil[*]{ortizgca@fisica.uaz.edu.mx}

\begin{abstract}
The least action principle played a central role in the development of modern physics. A major drawback of the principle is that its applicability is limited to holonomic constraints. In the present work, we investigate the energy lost by particles as a result of the gravitational interaction in a homogeneous low-density medium subject to non-holonomic constraints. We perform the calculation for an arbitrary particle and outline the specific result for photons. The energy lost is calculated from first principles based on the principle of virtual work and the d'Alembert principle. Under the formalism mentioned above, the dissipative nature of the effect is established. Furthermore, we show that the results agree with an alternative derivation based on continuum mechanics and the Euler-Cauchy stress principle. 
\end{abstract}
\begin{document}

\flushbottom
\maketitle
%
%
\thispagestyle{empty}


\section*{Introduction}

The foundation of classical mechanics was established by Isaac Newton in 1687 \cite{newton} when he formulated the laws of motion. Newton described the dynamics of a system by considering the forces acting on it, which are given by the product of mass and acceleration, also known as the rate of change of momentum.

However, the development of variational principles by Lagrange and Hamilton provided a more elegant and powerful framework to describe the dynamics of mechanical systems \cite{lagrange, hamilton}. They reformulated Newton's ideas in terms of a scalar quantity called the energy of the system. This approach offered several advantages, such as the ability to derive the equations of motion from a single mathematical framework and the insights gained by exploiting symmetries and conservation laws.

Nevertheless, variational methods have limitations when applied to systems with nonholonomic constraints \cite{variationallanczos1986}. Holonomic constraints are relations that depend solely on the coordinates and, possibly, on time. Variational methods, which rely on the principle of least action, struggle to handle non-holonomic constraints effectively.

To overcome this limitation, the d'Alembert principle emerges as a generalization of Newton's mechanics. By introducing an inertial force that counterbalances the impressed forces, the d'Alembert principle allows us to achieve equilibrium in an accelerated system. This principle effectively transforms an accelerated system into a static one, allowing us to recast the system in a moving reference frame to a static one, making it compatible with relativistic effects \cite{variationallanczos1986}.

One of the key advantages of the d'Alembert principle is its applicability to non-holonomic constraints. Non-holonomic constraints are relations that involve differentials of the coordinates, and possibly time. The d'Alembert principle provides a powerful tool for analyzing systems with such constraints, allowing for a more comprehensive treatment of mechanical systems.

In this work, we focus on calculating the loss of energy experienced by a particle that is constrained to move at a constant velocity through an electrically neutral low-density medium. Thus, we only
consider the classical gravitational force between particles and the medium.
To analyze this situation, we utilize the formalism of the d'Alembert principle.

First, we verify the equilibrium of the system employing the principle of virtual work. This principle provides a convenient approach to establish the conditions for equilibrium in systems subject to constraint forces. Once equilibrium is established, we proceed to calculate the work performed by the test particle as it moves through the medium.

Furthermore, we explore an alternative method based on the Euler-Cauchy stress principle to investigate the same effect. This approach provides a complementary perspective and allows for a comparison of results obtained using a different formalism.

By studying the loss of energy in this specific scenario with non-holonomic constraints, we aim to deepen our understanding of the d'Alembert principle and its application in analyzing the dynamics of constrained systems. The insights gained from this investigation contribute to the broader field of classical mechanics and provide valuable knowledge for various branches of physics and engineering.

\section*{Results}

The principle of virtual work states that, for any given mechanical system in equilibrium, it is necessary and sufficient that the total virtual work of all the forces involved vanishes,
\begin{equation}\label{workk}
    \delta \omega \equiv  \Sigma_{\nu=1}^{N} {\bf{F}_\nu^e \delta q_\nu} =0.
\end{equation}
By considering a reaction force created by the motion, force of inertia, d'Alembert generalized the principle of virtual work to dynamical systems.
 Thus,  any position of a system in motion may be regarded as an equilibrium position if we add the force of inertia, $ {\bf I_\nu}=-m_\nu{\bf a_\nu} $, to the impressed force, $\bf F_\nu$, acting on particle of mass $m_\nu$, \cite{gantmakher1970lectures} 
\begin{equation}
    \delta \omega= \Sigma_{\nu=1}^{N} ({\bf{F}_\nu+ {\bf I}_\nu) \delta q_\nu} =0.
\end{equation}
This means that the reaction forces determined by the d'Alembert principle in dynamical conditions are the same reaction forces of a static system \cite{variationallanczos1986}. 

For a free particle, the effective force, ${\bf{F}_\nu^e}={\bf{F}_\nu+ {\bf I}_\nu}$, is zero and, if subject to a given constraint, equal to the negative the force of reaction, that is ${\bf{F}_\nu^e}={\bf{F}_\nu+ {\bf I}_\nu=-\bf R_\nu}$. For ideal constraints, the  virtual work due to  the corresponding  forces of reaction is zero 
 \begin{equation}\label{const}
    \delta \omega= \Sigma_{\nu=1}^{N} {\bf{R}_\nu \delta q_\nu} =0.
\end{equation}
It follows that for ideal constraints, reaction forces must be perpendicular to the allowed virtual displacements.

\subsection*{Gravitational friction}
 For a particle in a homogeneous and isotropic medium, the contributions of all forces due to the medium are cancelled out.  For a static particle, the aforementioned condition can be sustained with the usual Newtonian mechanics, but this is not the case for a moving particle.
 
Under this phenomenon, the total work done by the particle is not done at the expense of its kinetic energy, as by definition the particle is limited to move at a constant velocity, and thus the constraint equation can be regarded as $ g({\bf r,v},t)= {\bf r-v}t=0$. The energy loss of the particle is due to the gravitational interaction with the medium. Since this involves non-holonomic constraints, we must rely on the principle of virtual work and the d'Alembert principle. 

In a dynamical  system,  the principle of virtual displacement requires that $\sum F_i=0$ be satisfied for any time,  $t$,  at a given position $r_i=r_{0i}$, where the velocity is set to zero,  $v_i=0$.  The aforementioned implies that the formulation is valid even if the displacement involves physically infinite velocities, as the actual motion of the body is not taken into account \cite{variationallanczos1986}. 
  
To calculate the work done by the particle on the medium, we partition the system into three regions, one of them given by the generated plane where the particle lies orthogonal to its displacement, $\partial {\Omega}=\{(r,x,\phi)\mid r=r_0\}$; another region above the plane, $\Omega_{+}=   \{(r,x,\phi)\mid r >r_0\}$, and the other below it,  $\Omega_{\_}=\{(r,x,\phi)\mid r<r_0\}$.



\begin{figure}[h]%
\centering
\includegraphics[width=0.9\textwidth]{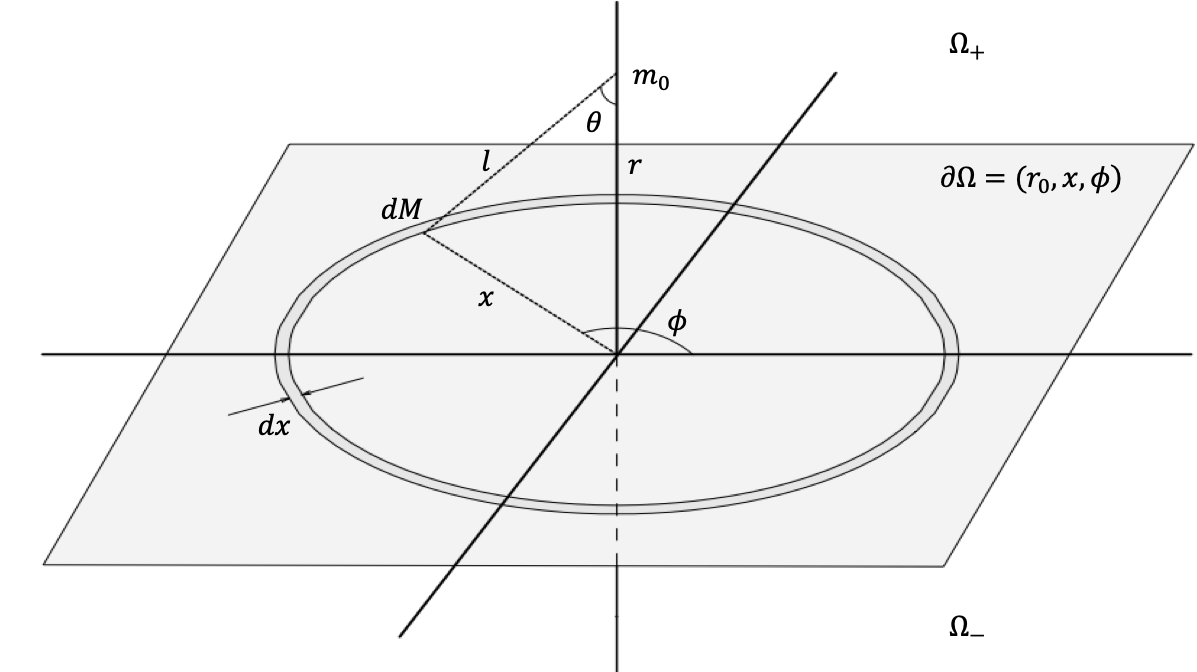}
\caption{Geometrical configuration of the system}\label{fig:plane}
\end{figure}

 Due to the symmetry of the configuration, the force on a particle due to the matter content in the upper region, $\Omega_{+}$, is canceled out by the force due to the one in the lower region, $ \Omega_{-}$, that is, $  F_{r^+}+F_{r^-}= $ $0$. Only the impressed  force ${\bf \bf F_\nu={F}}_{\partial {\Omega}}$ due the region in the plane $\partial {\Omega}$, orthogonal to the particle displacement, is left out. 
Given the constraint, $ g({\bf r,v},t)= {\bf r-v}t=0$, the acceleration vanishes, thus, the force of inertia disappears,  ${\bf I}_\nu=0$. So for any virtual displacement we have 
\begin{equation}
    \delta w= \Sigma {(\bf {F}}_{\partial \Omega})
    \cdot \delta{\bf r} = -  \Sigma {\bf R }\cdot
    \delta{\bf r}.
\end{equation}

The only forces involved are those due to the virtual displacement of the particle away from the original plane, $\partial \Omega$, and the reaction force $ \bf R$. 

The effective force is calculated for a virtual displacement along the $\hat{r}$ axis, as shown in Figure \ref{fig:plane}.  Due to symmetry, the sum of forces is zero in the $(x,\phi)$ plane,  $\Sigma  {\bf F}_{(x,\phi)}=0$, and  we only have the contribution of force in the $\hat{r}$ direction, which is orthogonal to the surface plane. Thus,  we calculate the force on a test particle $m_0$, located at a distance $r$ along the displacement, away from the plane of surface density $\sigma$. To do so, we calculate the force differential of the component $\hat{r}$ of concentric rings in the plane, with the differential mass $dM=\sigma 2 \pi x dx$,

   \begin{equation}
      dF_r=-Gm_0 \frac{2 \pi \sigma rx  d x}{(r^2+x^2)^{3/2}}.
  \end{equation}
The force on the particle due to the plane is given by the contributions of all the rings in the plane, that is, integrating from $x=0 \to x=\infty$, 
  \begin{equation} \label{fuerza}
      F_r=-2Gm_0 \pi \sigma. 
  \end{equation}
Since we are dealing with a virtual displacement away from the surface,  the surface density,  $\sigma$,  must be expressed in terms of the volume density, $\rho$,  $   M=\sigma A=\rho V$. Given the configuration of the displacement, see Figure \ref{fig:plane}, we assume the volume of a cone projected into a circle, hence $  \sigma =\rho r /3$. 

To obtain the force in the $\hat{r}$ direction,  we need to consider the constraints on the displacement in the $\hat{r}$ direction, $r=v_rt$, and substitute the superficial density in the force equation (\ref{fuerza}),
\begin{equation} \label{fuerzarho}
      F_r=-\frac{2}{3}\pi Gm_0\rho v_rt.
\end{equation}

 The displacement is along the variation, $\delta q= d r=v_r dt$,
\begin{equation}
   dW= -\frac{2}{3}\pi Gm_0\rho v_r ^2t  d t.
\end{equation}
We appreciate the explicit dependence on time, and thus it follows that   
 \begin{equation}
  \frac{ dW}{dt}= -\frac{2}{3}\pi Gm_0\rho v_r ^2t  \leq 0, 
\end{equation}
The previous equation is the condition of dissipativity  \cite{gantmakher1970lectures},  hence the connotation of friction in the proposed mechanism.

The total work for the displacement of a test particle $m_0$, at constant velocity, $v_z$, for an interval of time from $t_0=0$ to $t_f=t$, is given by 
\begin{equation} \label{work}
  W= -\frac{1}{3}\pi Gm_0\rho v_r ^2t^2. 
\end{equation}

   

It is worth noting that the energy-loss effect due to gravitational friction is independent of all other energy-loss mechanisms.  This is also a non-conservative system. Thus,  the energy is lost to the medium of concern.

  We acknowledge that while the d'Alembert principle is indeed well-suited for moving reference systems, it is not specifically formulated to handle curvilinear coordinates, particularly when considering general relativistic effects.To account for these effects, a more appropriate approach involves considering the variation of the Einstein-Hilbert action and treating the boundary terms as Lagrange multipliers. These Lagrange multipliers play a crucial role in incorporating non-holonomic constraints. Such  work is in progress, and it is considered to be published elsewhere.
\subsection*{Gravitational Surface Tension}
In the following, we apply the theory of continuous media \cite{Continuumchadwick} to calculate the potential energy due to surface tension which consequently gives the energy loss due to gravitational friction.


In continuous media theory, the Euler-Cauchy stress principle states that on any closed surface (real or imaginary) that divides a body, the action of one part of the body on the other is equivalent to its external forces acting on it, i.e., body forces $F_b$ and contact (Surface) forces or stress $F_s$ \cite{fung2001classical}.  
 
It is known that materials that manifest contact forces are composed of non-polar materials \cite{Continuumchadwick}. The gravitational potential which exhibits attractive force, has a non-polar behaviour. Therefore, as part of our hypothesis, we introduce the gravitational potential in the surface tension relation. Although the gravitational potential is several orders of magnitude weaker than the rest of the forces, we know that the gravitational potential is relevant at large distances.  We must acknowledge that the aforementioned hypothesis was previously developed in a general relativistic context \cite{carlos}, with a different setup.

In continuous media theory, the required work to increase a surface area, $A$, due to the surface tension is given by,

\begin{equation}\label{tension}
 \Delta   W_S=  \gamma_S \Delta A.
\end{equation}
We consider a half-sphere area differential $\Delta A=4\pi r dr$. The surface tension, due to the stress, is given by $\gamma_s(R)=\frac{F_S}{2R}$, where the coefficient of two in the denominator represents the two sides of the area of the object. The $R=2\pi r$ is the projection of half the sphere into the circle,

\begin{equation}
  \Delta W_S= \frac{F_s}{2\pi r}2\pi r \Delta r. 
\end{equation}

According to the preset work hypothesis, we consider  $F_S$  the force due to the gravitational potential.

\begin{equation}
\Delta  W_S=  -\frac{Gm_0M}{r^3} r \Delta r. 
\end{equation}
The mass is related to the density of the half-sphere, $M=(2/3)\pi \rho r^3$, thus, 
\begin{equation}
 \Delta  W_S= - \frac{2 Gm_0 \pi\rho}{3} r \Delta r. 
 \end{equation}

The surface tension potential energy, $PE_s$,  due to the gravitational potential is, 
\begin{equation}
    \Delta U =-W_S=\frac{1}{3} Gm_0 \pi \rho r^2.  \end{equation}\label{PE}


This result agrees with the expression for energy loss obtained by the virtual work framework (\ref{work}). This similarity comes at no surprise, since the development of the concept of stress is based on the principles of statics \cite{cauchy_2009}, i.e. equilibrium of forces and torques, which makes use of the postulate of virtual work \cite{stress_classical}, \cite{dellisola:hal-01226235}.

We acknowledge the fact that there can be more approaches to the problem.

\subsection*{Photons in low density medium}

Photons traveling in a homogeneous low-density medium provide an interesting scenario that aligns with the conditions discussed in the present work. In such a medium, photons move at a constant velocity, which is an example of a non-holonomic constraint. As mentioned in the introduction, variational methods are not well-suited for handling such constraints, making alternative approaches necessary.

To determine the energy loss experienced by traveling photons due to the gravitational friction mechanism, we employ the concept of equivalent plane waves. By considering the momentum of the photon, denoted as $p$, we can reframe Equation (\ref{work}) as follows:


\begin{equation} \label{w}
  W= -\frac{1}{3 c}\pi G p\rho r^2.   
\end{equation}



It must be pointed out that the energy of individual photons is lost, due to the work done by the medium opposing the photon's displacement. Since photons are constrained to move at a constant velocity, the loss of energy is elucidated as a change of momentum, resulting in the redshift of the photon's wavelength. The energy lost this way is released to the medium. The effect is complementary to other redshift mechanisms.

Given the dissipative nature of this mechanism, photons cannot gain energy through this process. As a result, any possibility of a blueshift arising from this mechanism is ruled out.


\section*{Discussion}

 At this point, let us discuss why this phenomenon has not been acknowledged before. We attribute it to the following factors, among others: first,  the effect is more perceptible in a density with a greater mean free path distance than the path displacement of the considered particle; otherwise, the effect is screened by other energy-loss mechanisms, such as scattering. 
The other factor is due to the nonholonomic constraints that are present in the system, which limit the applicability of the variational principle. The limitation is overcome by using the d'Alembert principle, which is a more general formalism.

We must acknowledge that the energy loss of photons due to gravitational friction can be easily differentiated from the usual scattering and diffraction under the considered non-holonomic constraint of constant-velocity. The energy loss under these effects involves a decrease in the number of photons or a path deviation, implying a blur on the image of distant objects. 
 
 
 Similarly, the gravitational energy loss, which is characterised by a conservative gravitational field, does not have a net effect for a complete closed path. However, in contrast to this, the gravitational friction mechanism has a dissipative effect for any distance covered by the particle.

 Considering the unique characteristics of gravitational friction in comparison to other energy-loss mechanisms, it becomes possible to devise an appropriate experimental setup for its measurement. One potential experiment that could satisfy the necessary conditions is the utilization of a gravitational interferometer such as the Laser Interferometer Gravitational-Wave Observatory (LIGO) \cite{Aasi_2015}. The suitability of LIGO as an experimental platform for investigating the gravitational friction phenomenon arises from its ability to meet several crucial criteria. Firstly, it provides a low-density environment, which is important for isolating the effects of gravitational friction from other energy loss mechanisms. Secondly, being located within the solar system, LIGO offers a convenient proximity for conducting experiments and making precise measurements. Lastly, its configuration includes a closed path loop for the observed photons.

LIGO operates in the near-infrared range, with a specific wavelength of 1064 nm. This fixed wavelength provides a constant value for the momentum term in the work equation (\ref{w}). By manipulating the density of the vacuum chambers, we can establish a linear relationship between the energy loss and the density. Additionally, the relationship between energy loss and distance can be tested by altering the number of cycles in the 4 km arm of the LIGO interferometer.

However, it is important to acknowledge that the feasibility of such an experiment may be limited by technical details that are currently unknown to us. It is crucial to consider the practical aspects and potential challenges associated with implementing the experimental setup at LIGO.

\section*{Conclusion}\label{sec13} 

In conclusion, our study reveals a novel mechanism by which particles lose energy through gravitational interaction with the medium they traverse. We have quantitatively determined the amount of energy lost by the particles as a function of the distance traveled and the average density of the medium. Our findings shed light on the dissipative nature of energy loss for particles moving at constant velocities, including photons.

The energy loss mechanism presented in this work is derived from first principles, specifically the d'Alembert principle. We have demonstrated that this mechanism aligns with an alternative derivation based on the Euler-Cauchy stress principle. The agreement between the two approaches provides additional confidence in the validity of our results.

Differentiating the gravitational friction energy-loss mechanism from other mechanisms is an important aspect discussed in our study. These distinctions provide valuable insights for designing experiments aimed at verifying the theory. 

Furthermore, our findings have significant implications for understanding the redshift of electromagnetic waves. By considering the energy loss of photons within this formalism, we can potentially explain a contribution to the observed redshift. The redshift of electromagnetic waves plays a crucial role in estimating physical quantities such as distance, velocity, and temperature. Thus, our research opens up new avenues for refining and expanding our understanding of cosmological observations.

In summary, our study has uncovered a gravitational friction energy-loss mechanism that affects particles moving through a medium. We have provided a quantitative analysis of the energy loss and established its dissipative nature. The application of the d'Alembert principle and its agreement with the Euler-Cauchy stress principle offer a robust foundation for our findings. We emphasize the importance of experimental verification and highlight the potential implications for explaining the redshift of electromagnetic waves. This research contributes to the advancement of our knowledge of gravitational interactions and their effects on particle dynamics in various physical systems.





\section*{Data availability}
The datasets used and/or analysed during the current study available from the corresponding author on reasonable request.
\bibliography{refe}


\section*{Acknowledgements }

CO acknowledges the support provided by project UAZ-2021-38486.

\section*{Author contributions statement}

All authors jointly interpreted the results and co-wrote the manuscript.  The Gravitational Friction representation was conceptualized by both authors and performed by C.O. The Surface Tension representation was developed by C.O. 

\section*{Additional information}

The corresponding author is responsible for submitting a \href{http://www.nature.com/srep/policies/index.html#competing}{competing interests statement} on behalf of all authors of the paper. 


\end{document}